\documentclass[useAMS,usenatbib,fleqn]{mnras}
\usepackage{newtxtext,newtxmath}
\usepackage[T1]{fontenc}
\usepackage{ae,aecompl}
\usepackage{graphicx}   % Including figure files
\usepackage{amsmath}    % Advanced maths commands
\usepackage{amssymb,times, multirow}    % Extra maths symbols

\def\pz{PZ~Mon}
\def\rs{RS~CVn}
\def\Teff{$T_{eff}$}
\def\Tspot{$T_{s}$}
\def\Twarm{$T_{w}$}
\def\Sspot{$S_{s}$}
\def\Swarm{$S_{w}$}

\def\lgg{log~$g$}
\def\kms{km~s$^{-1}$}
\def\vt{$v_{\rm t}$}
\def\deg{$^\circ$}

\def\jhk{\hbox{$J\!H\!K$}}
\def\vr{\hbox{$V\!-\!R$}}
\def\vi{\hbox{$V\!-\!I$}}
\def\vj{\hbox{$V\!-\!J$}}
\def\vh{\hbox{$V\!-\!H$}}
\def\vk{\hbox{$V\!-\!K$}}
\def\ebv{E(\bv)}
\def\ub{\hbox{$U\!-\!B$}}
\def\bv{\hbox{$B\!-\!V$}}

\begin{document}

\title[Spotted surface structure of the active giant \pz]{Spotted surface structure of the active giant \pz}

\author[Yu.V.Pakhomov et al.]{
Yu.~V.~Pakhomov$^{1}$,
K.~A.~Antonyuk$^{2}$, 
N.~I.~Bondar'$^{2}$, 
N.~V.~Pit'$^{2}$,
I.~V.~Reva$^{3}$, 
A.~V.~Kusakin$^{3}$
 \\
$^1$Institute of Astronomy, Russian Academy of Sciences, Pyatnitskaya 48, 119017, Moscow, Russia\\
$^2$Crimean Astrophysical Observatory RAS, Nauchny, Republic of Crimea, Russia
}

\date{\noindent
ISSN 1063-7737, DOI: 10.1134/S1063773717120040\\
Astronomy Letters, 2018, Vol. 44, No. 1, pp. 34--47.}

\pagerange{\pageref{firstpage}--\pageref{lastpage}}
\pubyear{}

\maketitle
\label{firstpage}

\begin{abstract}
Based on our photometric observations in 2015--2016 and archival photometric data for the active red giant \pz, we have found the main characteristics of the stellar surface: the unspotted surface temperature \Teff=4730~K, the spot temperature \Tspot=3500~K, and the relative spot area from 30 to 40\%. The best agreement with the observations has been achieved in our three-spot model including a cool polar spot with a temperature of $\sim$3500~K as well as large and small warm spots with a temperature of $\sim$4500~K. The stable polar spot is responsible for the long-period brightness variations. Its presence is confirmed by an analysis of the TiO 7054~\AA\ molecular band. The small-amplitude 34-day variability is attributable to the warm spots located on the side of the secondary component, which determine the relatively stable active longitude.
\end{abstract}

\begin{keywords} 
stars: individual: \pz\ --
(stars:) starspots --
stars: variables: general
\end{keywords}

\section{Introduction}

The active red giant \pz\ (HD 289114, $V\approx9$~mag, K2III) belongs to the class of \rs\ variables \citep{2015MNRAS.446...56P} whose brightness variations are associated with two main factors: a variable stellar activity manifested in the presence of large stable spots and a nonuniform distribution of smaller spots over the surface. The first factor is manifested in long-period brightness variations like the 11-year solar activity cycle. Periods of $\sim$ 60, 20, and 7 years were found for \pz\ \citep{2007OAP....20...14B}. The second factor is associated with rotational modulation and, therefore, the variability has a period equal to the stellar rotation period. A photometric period of $\sim$ 34~days was revealed by a periodogram analysis \citep{2007OAP....20...14B}, which was confirmed by \citep{2015MNRAS.446...56P} ? the measured rotational velocity of 10.5~\kms\ corresponds to a refined period of 34.13$\pm$0.02~days and an inclination of the rotation axis sin~$i$=0.92 for a red giant with a radius of 7.7~$R_{\sun}$ .

The spots cooler than the stellar photosphere are believed to be caused by a magnetic field that suppresses the convective energy transport. The magnetic field structure is visible on the Sun but is unobservable on stars. However, it determines the spotted stellar surface structure that manifests itself in brightness variability. Thus, an investigation of the structure and evolution of spots makes it possible to study the structure and evolution of stellar magnetic fields. In turn, different types of variable stars with spots reveal different factors affecting the magnetic field structure, among which are the rotational velocity, the evolutionary stage, the binarity, and others. 

In view of the relatively small amplitude of its brightness variations ($\Delta m_V\approx 0.05$), \pz\ has seldom become an object of photometric studies. The observations of the photometric ASAS-3 survey \citep{1997AcA....47..467P}\footnote{http://www.astrouw.edu.pl/asas/} performed in 2000?2009 are most complete. However, their accuracy of $\sim$0.02--0.04~mag, comparable to the brightness amplitude, is insufficient to investigate the surface brightness distribution. It can only be said as a first approximation that the active region has a significant size and always faces the secondary component \citep{2015AstL...41..677P}. The activity in \rs\ stars is induced by the influence of a close secondary component whose mass is comparable to the mass of the central star. However, in the case of \pz, we already see a manifestation of its activity in the presence of a low-mass companion (its mass ratio $M_2/M_1\approx0.09$ is minimal among all of such \rs\ objects), whose revolution, besides, is synchronous with the stellar rotation. From this point of view there is an interest in investigating the spotted surface structure of this object.

The first accurate long-term photometric observations of \pz\ (an accuracy of $\sim$0.007~mag) were performed in 2015 over slightly more than one stellar
rotation period \citep{2017Ap.....60..395A}. These suggested a relative stability of the active longitude over the last 15 years \citep{2017ASPC..510..128P}. The authors explain the slight phase shift of the light curve relative to the ephemeris either by an inaccurate period or by differential rotation and a nonuniform distribution of spots in the active region. 

The goal of this paper is to determine the surface characteristics of \pz\ by analyzing our new 2015--2016 photometric data. In the first section we describe our photometric observations. In the next section we also use archival infrared photometric data and spectroscopic observations near the TiO 7054~\AA\ molecular band to estimate the stellar spottedness and the spot temperature. This enables us to construct a model for the distribution of spots on the surface of \pz\ that describes its photometric characteristics.

\section{OBSERVATIONS}

% Table 1
\begin{table} 
\renewcommand{\tabcolsep}{0.6mm}
\caption{Photometric characteristics of \pz.\label{tab:chars}}
\centering
\begin{tabular}{|l|c|c|c|c|c|}
\hline
\multicolumn{1}{|c|}{Parameter}                   & $U$ & $B$ & $V$ & $R$ & $I$ \\
\hline
&\multicolumn{5}{|c|}{Epoch 1}\\
\cline{2-6}\\
Range, mag &       --      & 10.28-10.40 & 9.11-9.22 & 8.48-8.58 & -- \\
Amplitude, mag          &       --      &  0.12$\pm$0.01&0.11$\pm$0.01&0.10$\pm$0.02& -- \\ 
Spots area $S$, \%       &       --      &  31.7$\pm$0.7 &30.4$\pm$0.7 &29.2$\pm$1.4 & -- \\
\hline
&\multicolumn{5}{|c|}{Epoch 2} \\
\cline{2-6}\\
Range, mag & 11.30-11.47 & 10.39-10.55 & 9.24-9.37 & 8.62-8.70 & 8.01-8.11 \\
Amplitude, mag          &  0.17$\pm$0.04&  0.14$\pm$0.02&0.13$\pm$0.02&0.08$\pm$0.03& 0.10$\pm$0.04  \\ 
Spots area $S$, \%   &  40.1$\pm$2.4 &  38.7$\pm$1.2 &38.8$\pm$1.2 &38.8$\pm$1.3 & 37.9$\pm$2.4 \\
\hline
\end{tabular}
\end{table}

Photometric observations of \pz\ were carried out in two periods that we will call epochs below. From January 23 to April 5, 2015, (epoch~1) the observations were carried out with the 1.25-m AZT-11 telescope of the Crimean Astrophysical Observatory (CrAO) using a CCD photometer. We performed 21 observations in three bands, BVR$_C$ , each of which consisted of several 180-s exposures (a total of 94 frames). The total time of our observations covers about two orbital periods of \pz. The accuracy of our magnitude measurements ranged from 0.007 to 0.01~mag. Simultaneously with the photometric observations, from January 18 to 27, 2015, we carried spectroscopic observations of \pz\ with
the MAESTRO echelle spectrograph (R = 40\,000) mounted at the Coude-focus of the 2-m Zeiss-2000 telescope of the Terskol branch of the Institute of Astronomy, the Russian Academy of Sciences (Peak Terskol, Elbrus). The spectra were reduced in the MIDAS package; 88 orders cover the range from 3530 to
10065~\AA.

% Figure 1
\begin{figure}
\centering
\includegraphics[width=0.45\textwidth,clip]{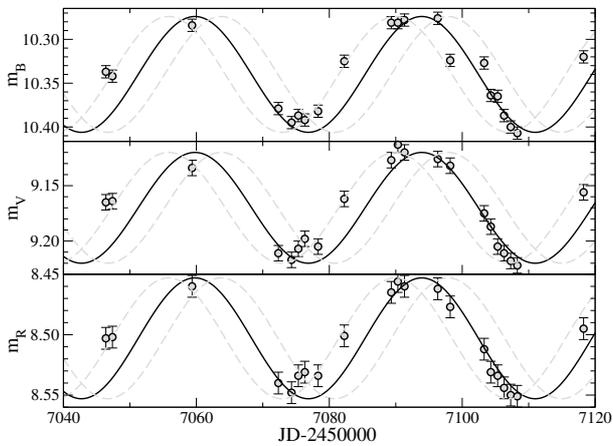}
\caption{ $BVR$ light curves of \pz\ from the CrAO observations (epoch~1). The thick line indicates the ephemeris with a 3$\sigma$ error in the period (dashed grey line).}
\label{fig:Crao}
\end{figure}

% Figure 2
\begin{figure}
\centering
\includegraphics[width=0.45\textwidth,clip]{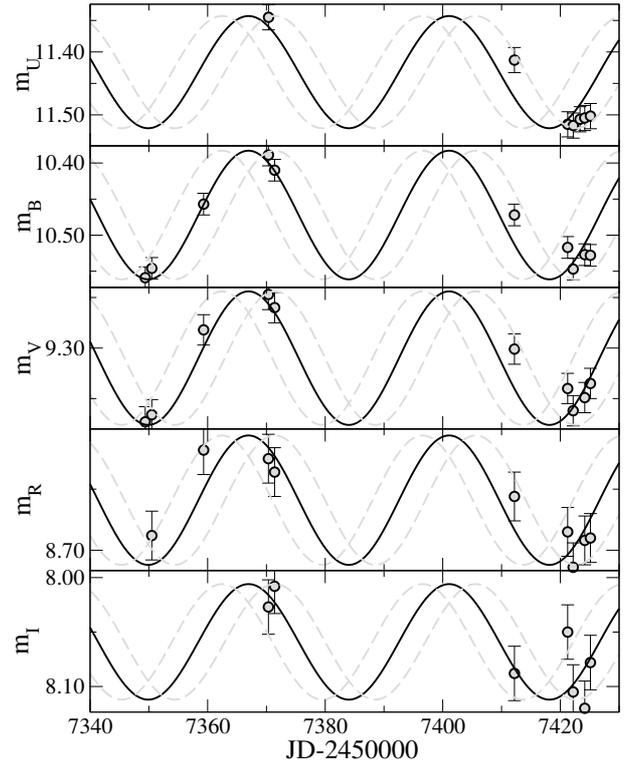}
\caption{$UBVRI$ light curves of \pz\ from the observations at the Tian Shan Observatory (epoch~2). The thick line indicates the ephemeris with a 3$\sigma$ error in the period (dashed grey line).}
\label{fig:AA}
\end{figure}

% Figure 3
\begin{figure}
\centering
\includegraphics[width=0.5\textwidth,clip]{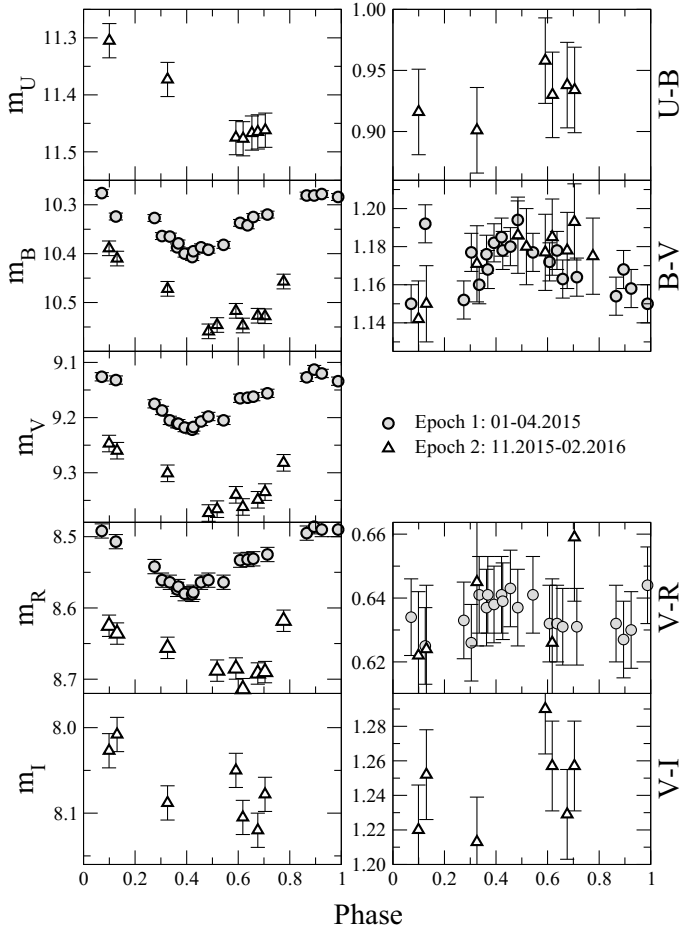}
\caption{ Light and color curves of \pz\ folded with the 34.13~day period. The circles and triangles indicate the data for
epochs~1 and 2, respectively.}
\label{fig:all}
\end{figure} 

From November 22 to December 2, 2015, and from December 13, 2015, to February 6, 2016, (epoch~2) the observations were carried out, respectively, with the Western and Eastern Zeiss-1000 telescopes at the Tian Shan Astronomical Observatory of the Fesenkov Astrophysical Institute using a CCD photometer. Over 10 nights we took 454 CCD frames with an exposure time from 10 to 180~s in
five bands, UBVR$_C$I$_C$ . We reduced the magnitudes to the standard Johnson?Cousins photometric system using stars in the frame based on data from the APASS catalog \citep{2015AAS...22533616H}. The accuracy of our magnitude measurements ranged from 0.015 to 0.025~mag. In Figs.~\ref{fig:Crao} and \ref{fig:AA} the circles with error bars indicate the light curves for the observations of epochs~1 and 2, respectively. The thick line draws The ephemeris for maximum light $m=\overline{m}-\Delta{m}\, \textrm{cos}(2\pi(JD-2454807.2)/P)$, where $P$=34.13~days is the previously determined photometric period \citep{2015MNRAS.446...56P}, is drawn by the thick line. The amplitudes $\Delta{m}$ and mean magnitudes $\overline{m}$ were chosen for each set of observations. The grey dashed line marks the ephemerides with 3$\sigma$ errors in the period. It is clearly seen that at the first epoch the observed light curve slightly leads the ephemeris, while at the second one, on the contrary, it slightly lags behind. Fig.~\ref{fig:all} shows the light and color curves folded with the period. Over 2015 \pz\ faded approximately by 0.10--0.15~mag. The photometric characteristics of \pz\ are reflected in Table~\ref{tab:chars}, where the range of brightness variations in the bands, the brightness amplitudes, and the relative spot area corresponding to maximum light (see below) are given. For details on the photometry of \pz\ at epoch~1, see \cite{2017Ap.....60..395A}.

\section{PHOTOMETRIC CHARACTERISTICS OF \pz}

The observational data obtained in 2015--2016 show the following photometric characteristics of the star under study: a decrease in the mean brightness over 2015, while the behavior of the colors is unchanged; periodic brightness variations in various photometric bands, with the amplitude decreasing with increasing wavelength; and periodic color variations. In addition, archival broadband photometry reveals that different colors describe different effective temperatures of the star. For example, on JD 2451498 the 2MASS catalog \citep{2003yCat.2246....0C} presents the infrared JHK magnitudes of \pz, the $UBVRI$ magnitudes on the same date can be estimated from \cite{2006A&AT...25..247A}. Whereas the \ub\ and \bv\ colors correspond to
the model with parameters \Teff=4700~K, \lgg=2.8, [Fe/H]=0.07, and \ebv=0.06~mag determined by \cite{2015MNRAS.446...56P}, \vr\ and \vi\ are slightly increased by 0.05--0.14~mag relative to this model, and such values are typical for stars with \Teff=4510 and 4400~K. The infrared \vj, \vh, and \vk\ colors already differ significantly from the model ones by 0.31~mag , 0.35~mag, and 0.41~mag, respectively,typical for stars with \Teff=4360--4460~K. Such a picture is observed for spotted stars. Indeed, at a lower spot temperature ($\Delta T\approx 1000-1500$~K) the intensity of stellar surface radiation in the blue and yellow parts of the spectrum ($UBV$) dominates
considerably over the intensity of spot radiation: $F_U^{4700K}/F_U^{3500K}\approx50$, $F_B^{4700K}/F_B^{3500K}\approx20$, and $F_V^{4700K}/F_V^{3500K}\approx15$ , while in the infrared \jhk\ bands the ratio $F^{4700K}/F^{3500K}$ is from 1.5 to 3, and the spot radiation is already noticeable. Thus, the spectral energy distribution for \pz\ cannot be described by a model stellar atmosphere with a single temperature.

% Figure 4
\begin{figure}
\centering
\includegraphics[width=0.48\textwidth,clip]{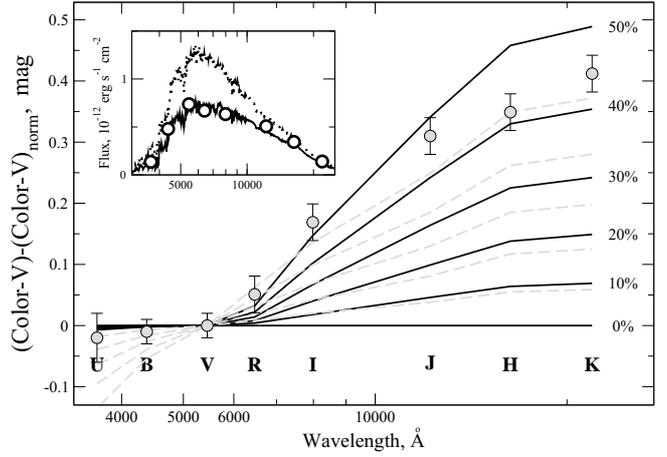}
\caption{
Color difference from a normal star with parameters \Teff/\lgg/[Fe/H]/\ebv = 4700/2.8/0.07/0.06. The circles mark the observed values for \pz\ for the epoch JD 2451498, the thick lines indicate the theoretical values for spots at
\Tspot=3500~K with a relative area S from 0 to 50\%, and the grey lines indicate the same theoretical values for \Tspot=4000~K. The inset shows the spectral energy distributions for \pz\ (circles), a normal star (dotted line), and a star with spots at \Tspot=3500~K and $S$=45\% (solid line).}
\label{fig:dm}
\end{figure}

Figure~\ref{fig:dm} shows the observed color excesses for \pz\ relative to the calculated values for a star with parameters \Teff=4700~K, \lgg=2.8, [Fe/H]=0.07, and the interstellar color excess \ebv=0.06~mag as well as the calculated color excesses for a spotted star with a relative area of spots $S$ with a temperature
of 3500~K from 0 to 50\%. The calculations were performed by interpolating the grid of magnitudes $m_X$ (\Teff, \lgg, [Fe/H]) in the corresponding bands $X$
from \cite{1998A&A...333..231B} \footnote{http://wwwuser.oats.inaf.it/castelli/grids.html}. The magnitudes of the spotted star were calculated with a partial allowance made for the limb darkening (because the distribution of spots over the surface is not considered, but the magnitudes mX themselves from the grid of models take this effect into account) from the relation

\begin{eqnarray}
m_X = -2.5\,lg \left(10^{-0.4 \displaystyle m_X^s}\,S+
10^{-0.4 \displaystyle m_X}\,(1-S) \right) -
\nonumber\\
-5\, lg\frac{\theta}{2} + E(B-V) R
\label{eq:m}
\end{eqnarray}

where $m^s_X$ is the magnitude of the starspot at \Tspot=3500~K, $m_X$ is the magnitude of the unspotted stellar surface at \Teff=4700~K, $\theta$=0.29~mas is the angular diameter of the star, and $R=A/E(B-V)$ is the ratio of the extinction in a given band to the color excess \bv. For the $UBVRI$ bands we took the following values of R: 4.78, 4.08, 3.10, 2.58, and 1.85.

It follows from Fig.~\ref{fig:dm} that the behaviour of the observed colors for \pz\ corresponds to a star with approximately 45\% of its surface covered with cool spots. The observed and calculated spectral energy distributions for \pz\ are shown in the inset of Fig.~\ref{fig:dm}, where the spectrum of a normal red giant with similar parameters is also presented for comparison. For clarity, Fig.~\ref{fig:dm} also shows the calculations for a spot temperature of 4000~K. In this case, \ub\ and \bv\ will be sensitive to the spot area, which is not observed in \pz; consequently, the spot temperature is below 4000~K. Hence it also follows that the behaviour of spectral lines in the optical ($V$) band reflects the characteristics of precisely the unspotted stellar surface. Therefore, the atmospheric parameters of \pz\ determined by \cite{2015MNRAS.446...56P} correspond to a quiet photosphere. The brightness of \pz\ in the periods of maximum light, when the stellar activity and the spot area are minimal, must
correspond to such parameters. The theoretical $B$ magnitude with the above parameters and an angular diameter of 0.29~mas is 9.89~mag . According to \cite{2006A&AT...25..247A}, the maximum brightness in 1938 and 1983 reached mph — 9$^m$.94, corresponding to $m_B = m_{ph} + 0.05 = 9^m.99$ \citep{1961ApJ...133..869A}. This is consistent, within the measurement accuracy, with the theoretical values of the photographic magnitude and the possible presence of a few spots even in the periods of maximum light. At this time, the relative spot area $S$ on the surface of \pz\ (from the relation $\Delta\,m = -2.5 lg(1-S(1-I_0/I_{spot}))$, where
$I_0/I_{spot}\approx 1/20$ is the ratio of the intensities of radiation from the unspotted surface and the spots in the $B$ band) reached 9\%.

Such an approach of comparing the observed magnitudes of \pz\ at the times of maximum light with the calculated ones can be used to estimate the relative spot area. Table~\ref{tab:chars} gives the values of $S$ for spots with a temperature of 3500~K determined for each of the $UBVRI$ bands. These values are seen to agree well between themselves.

\subsection{Spot characteristics from the observations at Epoch 2}

Since the Tian Shan data are presented in a wider range, it is possible to estimate not only the relative surface area of the spots but also their temperature from Eq.~(\ref{eq:m}) using the Levenberg--Marquardt non-linear fitting method. The best agreement with the observed maximum values of the magnitudes in different bands is achieved at \Teff=4716$\pm$20~K, \Tspot=3500$\pm$300~K, and $S$=41.2$\pm$0.7\%. The large error in the spot temperature suggests a weak sensitivity of this parameter, and any value below 3800~K describes the observed magnitudes almost equally well, within the measurement error limits. 

Thus, the brightness variability of the star \pz\ and its spectral energy distribution can be explained qualitatively and quantitatively by the presence of cool spots. However, the two-temperature model does not reflect all features of the brightness variations in \pz. For example, it cannot describe the observed change in the \bv\ color with the stellar rotation period. This requires that the spot temperature be much closer to the surface temperature of \pz. The three-temperature model with the addition of warm spots, by analogy with the penumbras of active regions on the Sun, appears more realistic. Modelling of \rs\ stars also shows that the active regions are characterized by a wide spread in temperatures \citep[see, e.g.][]{2002A&A...389..202O}. In this case, the observed magnitudes are described by the formula

\begin{eqnarray}
m_X = -2.5\,lg [10^{-0.4 \displaystyle m_X^s}\,S_s+10^{-0.4 \displaystyle m_X^w}\,S_w + 
\nonumber\\
+ 10^{-0.4 \displaystyle m_X}\,(1-S_s-S_w)] -
\nonumber\\-5\, lg\frac{\theta}{2} + E(B-V) R
\label{eq:mpen}
\end{eqnarray}
where the indices $s$ and $w$ refer to the cool and warm spots, respectively, and the absence of an index refers to the unspotted surface. This expression differs from~(\ref{eq:m}) by the addition of a parameter that characterizes the warm spots by the area \Swarm. In this case, the Levenberg--Marquardt method gives \Teff=4747$\pm$90~K, \Twarm=4490$\pm$210~K, \Sspot=39.3$\pm$0.7\%, \Swarm$=14\pm12$\%. For these parameters the observed magnitudes are described with a 0.011~mag accuracy. Compared to the two-temperature model, the stellar surface temperature increases insignificantly, within the error limits, while the area occupied by the cool spots remains the same. A difference \Teff-\Twarm$\approx$250~K corresponds to solar penumbras; the area of these warm spots is considerably smaller than that of the cool ones. However, their characteristics have large uncertainties, and, within the limits of the errors
in the observed magnitudes, we can abandon the presence of warm spots for the case of maximum light altogether. At minimum light the \bv\ color index increases from 1.155 to 1.182, which, given the reddening \ebv=0.06, corresponds to a decrease in the mean temperature from 4730 to 4650~K. The former characterizes the unspotted part of the star, while the latter reflects the presence of warm spots. We obtain the parameters \Teff=4770$\pm$90~K, \Twarm=4200$\pm$500~K, \Sspot=38$\pm$2\%, \Swarm$= 24\pm 12$\% by the optimization method. The area of the cool spots again remains constant, while that of the warm ones increases noticeably. Thus, the assumption that the 34-day variability during which the \bv\ color changes is caused by a warm spot is confirmed.

\subsection{Spot characteristics from the observations at Epoch 1}

Similar estimates of the spot characteristics were also made for the more accurate CrAO photometric data. Here, however, we are restricted to the data only in three bands. Therefore, to estimate the parameters of the spots, we will have to fix their temperatures \Tspot=3500~K, \Twarm=4500~K, \Teff=4730~K. For maximum light \Sspot=$34.6\pm1.5$\%, \Swarm$=0^{+2}_{-0}$\%.
We observe the same picture as above: to describe the magnitudes at maximum light, we can use the two-temperature model without invoking any warm spots. For minimum light the parameters were \Sspot=$33.0\pm0.3$\%, \Swarm=$27.6\pm1.0$\%. Thus, the area of the cool spots remains as before, while the brightness and color variations are provided by the warm spot.
At these parameters the observed magnitudes are described with a 0.009~mag accuracy. Thus, the photometric behavior of \pz\ in 2015--2016 can be characterized by a constantly visible cool spot that slowly changes in sizes and limits the maximum brightness in this period and a warm spot that is responsible for the brightness and color variability. The cool spot is most likely polar, because the brightness of the star does not change as it rotates. A similar picture is observed for \rs\ stars \citep{2002A&A...389..202O} and will be discussed below.

\subsection{The cool spot temperature and the TiO 7054~\AA\ band}

\begin{figure}
\centering
\includegraphics[width=0.45\textwidth,clip]{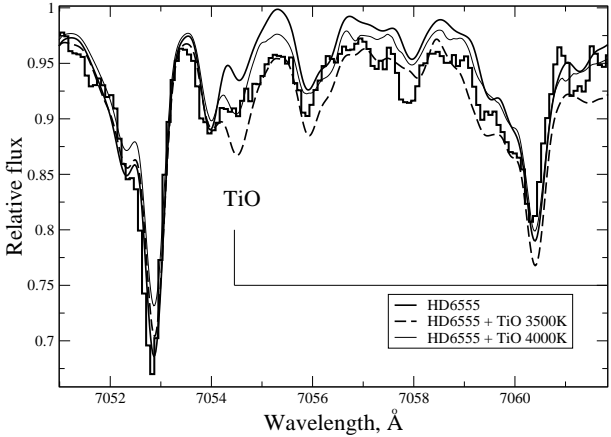}
\caption{The spectrum of \pz\ (histogram) in comparison with the spectrum of the normal red giant HD~6555 (thick line) and with the addition of the synthetic TiO molecular spectrum computed at $S=$30\%\ and \Tspot=3500~K (dashed line) and 4000~K (thin line).}
\label{fig:TiO}
\end{figure}

TiO molecular bands are a very sensitive indicator of the presence of cool spots with a temperature below 4000 K \citep{2002A&A...381..517A}. No TiO bands
have been revealed previously in the visible part of the spectrum, because in this range the flux from the unspotted part of the star exceeds considerably
the flux in its cool spots. Longer-wavelength ranges should be used in the search for TiO bands. The infrared region of the spectrum is significantly distorted by interference fringes formed in CCD layers and by strong telluric lines and, therefore, is unsuitable for analysis. The strongest TiO band (the $\gamma$ band, the $A^3\Phi $--$ X^3\Delta$ transition) in the red part of the spectrum, which is unaffected by fringes and telluric lines, begins abruptly at a wavelength of 7054.2~\AA\ and is detected in the spectrum of \pz. Since no variability of the molecular line intensity were detected within the noise limits in the time of our observations spanning almost a third of the stellar rotation period, 17 spectra were added. The signal-to-noise ratio in this interval was $\sim$80. Isolated telluric lines were taken into account using the spectrum of a star with a high rotational velocity. Figure 5 shows a portion of the spectrum for \pz\ near 7054~\AA; the spectrum of the normal red giant HD~6555 without any evidence of activity is presented for comparison.
We chose this star based on a similarity of the stellar atmosphere parameters (\Teff/\lgg/[Fe/H]/\vt = 4720/3.00/-0.09/1.15 \cite{2013AstL...39...54P}) and apparently close carbon and nitrogen abundances, which ensures the reproduction of CN molecular bands. The latter fact is important, because the beginning of the TiO band coincides with one of the CN lines. Furthermore, the spectrum of HD~6555 almost reaches the continuum level, suggesting the absence of TiO lines. Thus, the differences between the spectra in this range are expected to be associated with molecular TiO lines. Indeed, differences appear at wavelengths longer than 7054~\AA; to estimate them, we performed computations with the SYNTHV code and using the ATLAS9 stellar model atmospheres (\Teff=3500~K, 3750~K, and 4000~K), the molecular parameters of the TiO lines were taken from the VALD \citep{2015PhyS...90e4005R}. The computed molecular spectra were convolved with the instrumental profile ($R$=40\,000) and the rotation (10.5~\kms) and the microturbulence (6.5~\kms) profiles and were then summed with the spectrum of the comparison star HD~6555. The final spectrum was computed from the relation
$$
F_{res}=\frac{F^{TiO}_{res} R S + F^{star}_{res} (1-S)}{R S + (1-S)}
$$
where  $F^{TiO}_{res}$ is the residual flux for the TiO spectrum, $F^{star}_{res}$ is the residual flux for the spectrum of the comparison star HD~6555, $R$ is the ratio of the fluxes for these two spectra in continuum calculated from the model stellar atmospheres, and $S$ is the relative spot area that was taken to be 0.3 from an analysis of the photometric observations. The results are shown in Fig.~\ref{fig:TiO}. Because of some uncertainty in drawing the continuum level and the blending of the beginning of the TiO band by CN molecular lines, an accurate temperature of the cool spot is difficult to obtain for the spectrum of \pz. Different regions show agreement with the observations for Ts between 3500 and 4000~K; besides, the uncertainty in the relative spot area and in the molecular line parameters remains.

% Figure 5

\section{THE MODEL OF THE SPOTTED STRUCTURE OF \pz}

Any modelling of the intensity distribution over the stellar surface must describe the behaviour of the light curves in different colors, i.e., both brightness and color variations. Inverse methods are commonly used for modelling: the method of Doppler tomography if spectra with a high signal-to-noise ratio are available \citep[see, e.g.][]{2000A&A...357..608S, 2002A&A...389..202O} or the method of light curve reconstruction requiring a
dense series of highly accurate observations \citep[see, e.g.][]{1998A&A...332..541L, 2008AN....329..364S}. These methods are mathematically ill-defined, but they yield a result whose quality depends strongly on the accuracy of the input data when using a number of restrictions (Tikhonov?s regularization or the entropy maximum condition). The first method gives a more or less realistic picture; the second method is sensitive only to the spot longitude distribution. In the absence of simultaneous spectroscopic observations for \pz\ and given the low accuracy of the constructed light curves, parametric models can be used to describe the surface characteristics of a spotted star \citep{2002A&A...381..517A}. The sizes, positions, and temperatures of the spots whose number is chosen to be the minimum possible one serve as the model parameters. An advantage of this approach is the unique solution within the model being applied. The observational errors are transformed into the errors in parameters. The Levenberg--Marquardt method, which finds the local minimum of the O--C residuals, is used most commonly to calculate the parameters. Therefore, it is important to properly choose the initial values of the parameters.
 
\subsection{The technique of calculations}

To calculate the magnitudes, we produce a surface map in a rectangular coordinate system of equirectangular projection ($\lambda$, $\phi$) for the bands $X=U,B,V,R,I$ with sizes of 720x360 (0.5\deg\ step), which is then transformed into a sphere with inclination of 67\deg\ in an orthographic projection.coordinates ($\lambda$, $\phi$) are transformed into the spherical ones ($l$, $\psi$) describing the observed stellar hemisphere.

In each iteration step, based on the model parameters, we construct a mask specifying the surface type at each point: an unspotted one, a cool spot, a warm
spot. The intensity IX in band X at a fixed temperature was reconstructed by the method of interpolation from the grids of magnitudes $m_X(T_{eff},\textrm{lg}\,g, [Fe/H])$ (\cite{1998A&A...333..231B} \footnote{http://wwwuser.oats.inaf.it/castelli/colors/bcp.html}) calculated from the fluxes for 1 cm$^2$ of the stellar surface:
$$
I_X=\frac{10^{\displaystyle-0.4\,m_X}}{\pi(1-\frac{\displaystyle\epsilon_X}{\displaystyle 3})}
$$ 

where $\epsilon_X$ is the limb-darkening coefficient in band $X$ calculated by the interpolation method based on the data from \cite{2013A&A...554A..98N} for the plane-parallel ATLAS9 models that we also use.

The flux and magnitude in band $X$ are then

$$
F_X=\int_{0}^{2\pi}\int_{0}^{\pi/2} I_X ((1-\epsilon_X)+\epsilon_X cos(\psi)) \textrm{sin}(\psi) \textrm{cos}(\psi) \textrm{d}\psi \textrm{d}l 
$$
$$
m_X=-2.5\,lg F_X -5\, lg\frac{\theta}{2} + E(B-V)\cdot R_X
$$ 

The calculations were performed for the phases corresponding to the observations of \pz, and the difference of the observed and calculated magnitudes is minimized.

\subsection{The polar cool spot}

The above analysis of the photometric data has shown that cool spots are constantly present on the surface of \pz. Such a picture may be indicative
either of their equilibrium distribution or a concentration of some of them at the pole, as confirmed by the Doppler mapping of \rs\ stars \citep{2002A&A...389..202O, 2003A&A...402.1073G, 2000A&A...357..608S, 1998A&A...330.1029W, 2010A&A...515A..14K}. Therefore, in our model we place the spot at the pole. 

The basic model parameters for the surface of \pz\ will be the temperature of the unspotted region \Teff, the radius of the cool polar spot $R_{pol}$, and its temperature \Tspot. These parameters describe the brightness of \pz\ at its maximum, i.e., at phases close to zero. The first parameter is well determined from spectroscopic observations (\Teff=4700$\pm$100~K) and from the \bv\ color, whose minimum of 1.155~mag corresponds to \Teff=4730~K and reddening \ebv=0.06. The last two parameters
depend on each other; $R_{pol}$ increases with rising \Tspot. Therefore, we fix \Tspot=3500~K that corresponds best to the photometric data. 

As a result of our calculations, the radius of the polar spot is $R_{pol}$=60\deg\ for the observations of epoch~1 and 64\deg\ for epoch~2. Note that its minimum radius was 43\deg\ in 1938 and 1983, when \pz\ was at maximum light. Below we will fix these derived parameters, because there is no modulation with the stellar rotation.

% Table 2
\begin{table}  
\renewcommand{\tabcolsep}{2.5mm}
\caption{Parameters of the spots on the surface of \pz.\label{tab:param}}
\centering
\begin{tabular}{|l|l|c|c|}
\hline
Spot & Parameter                   & Epoch 1  & Epoch 2 \\
\hline
\multirow{2}{*}{Polar spot} & $R$, \deg  & 60$\pm$1 & 64$\pm$1 \\

                                & \Tspot, K  & 3500     & 3500     \\
\hline
\multirow{4}{*}{Main spot} & $l_0$, \deg& 29$\pm$3 & -6$\pm5$ \\

                                & $a_1$, \deg& 59$\pm$5 & 79$\pm$7 \\

                                & $a_2$, \deg& 65$\pm$5 & 106$\pm$7 \\

                                & $b$, \deg  & 10$\pm$1 & 9$\pm$2 \\

                                & \Twarm, K  & 4475$\pm$60 & 4428$\pm$70 \\
\hline
\multirow{2}{*}{Second spot}   & $l_2$, \deg&-62$\pm$7 & 119$\pm$20 \\

                                & $R_2$, \deg& 10$\pm$2 & 5$\pm$5 \\
\hline                               
\end{tabular}
\end{table} 

\subsection{The warm spots}

As has been shown above, the presence of warm spots with a temperature of $\sim$4500~K provides color variations and can explain the decrease in the amplitude of brightness variations in different bands from the blue region to the red one. To simplify the problem, let us assume that all warm spots are located on the equator, which is often observed in \rs\ stars \citep{2002A&A...389..202O}. The main spot responsible for the behaviour of the light curve must have a longitudinally elongated shape, with the degree of elongation differing in the western and eastern directions, as follows from the light curves. Therefore, we will take a spot in the shape of two halves of ellipses: $(\lambda-\lambda_0)^2/a^2+\phi^2/b^2=1$, where  $\lambda_0$ is the longitude of the center of the main spot, $b$ is the semiminor axis -- the width of the spot in latitude symmetrically to the equator, $a$ is the semimajor axis $a=a_1$ at $\lambda<\lambda_0$ and $a=a_2$ at $\lambda>\lambda_0$. We will take the initial meridian to be at phase 0.5, corresponding to the brightness minimum of \pz. Since the first derivative of
the light curve is nonmonotonic, we will add a second spot $(\lambda-\lambda_2)^2+\phi^2=R_2^2$ in the shape of a circle with radius $R_2$ and its center ($\lambda_2$, $\phi=0^\circ$). Thus, we have a total of seven parameters: six geometrical ones and one physical parameter -- the temperature taken to be constant for all warm spots.

Let us first estimate approximate values for the parameters. For epoch~1 the following values are taken to be the initial approximation: $\lambda_0$=36\deg, because the brightness minimum is located at a phase of about 0.40, and $\lambda_2$=54\deg, corresponding to phase 0.65, where a plateau is clearly seen on the light curve. Since the main spot itself occupies about 0.6 of the
visible surface, we will take $a_1=a_2=180\cdot0.6/2=54^\circ$. From our analysis of the photometric data we obtained the relative area of the warm spots, 27\%, whence $b$ = 15\deg, while for $R_2$ we will take 10\deg. The
initial temperature of the spots is \Twarm=4500~K. For epoch~2 we took the following values of the initial parameters:$\lambda_0$=-18\deg, $\lambda_2$=90\deg, $a_1=a_2=70^\circ$, $b=15^\circ$, $R_2=10^\circ$, \Twarm=4500~K. When bringing the calculated and observer light curves  closer, we used only the $B$ and $V$ data, because the absolute accuracy of the photometry in this case was much higher than that in other bands. 

\begin{figure}
\centering
\includegraphics[width=0.5\textwidth,clip]{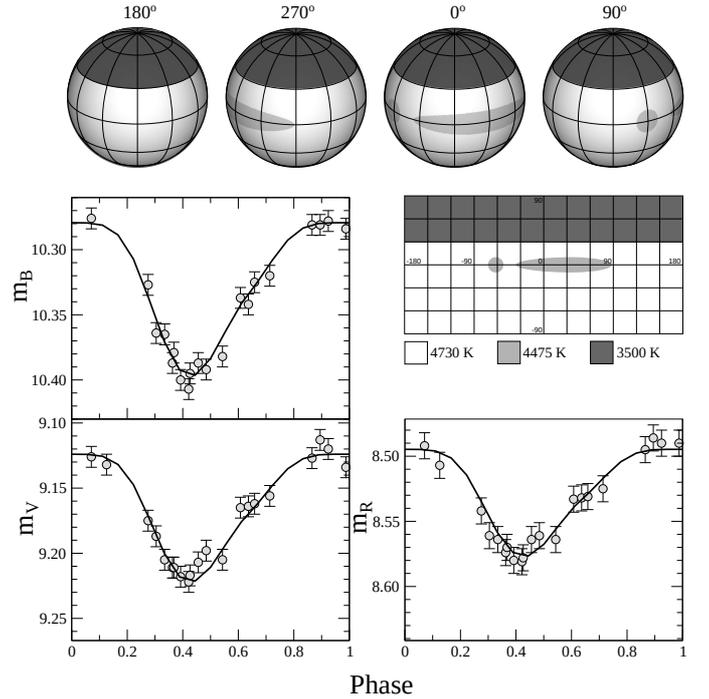}
\caption{Results of our modelling of the light curves and the surface of \pz\ from the observations at epoch~1 from January~23 to April~5, 2015. The view of the surface at different phases is at the top. The surface map is in the middle, and the remaining panels show the observed and calculated light curves in different bands.}
\label{fig:synth_Crao}
\end{figure}

\begin{figure}
\centering
\includegraphics[width=0.5\textwidth,clip]{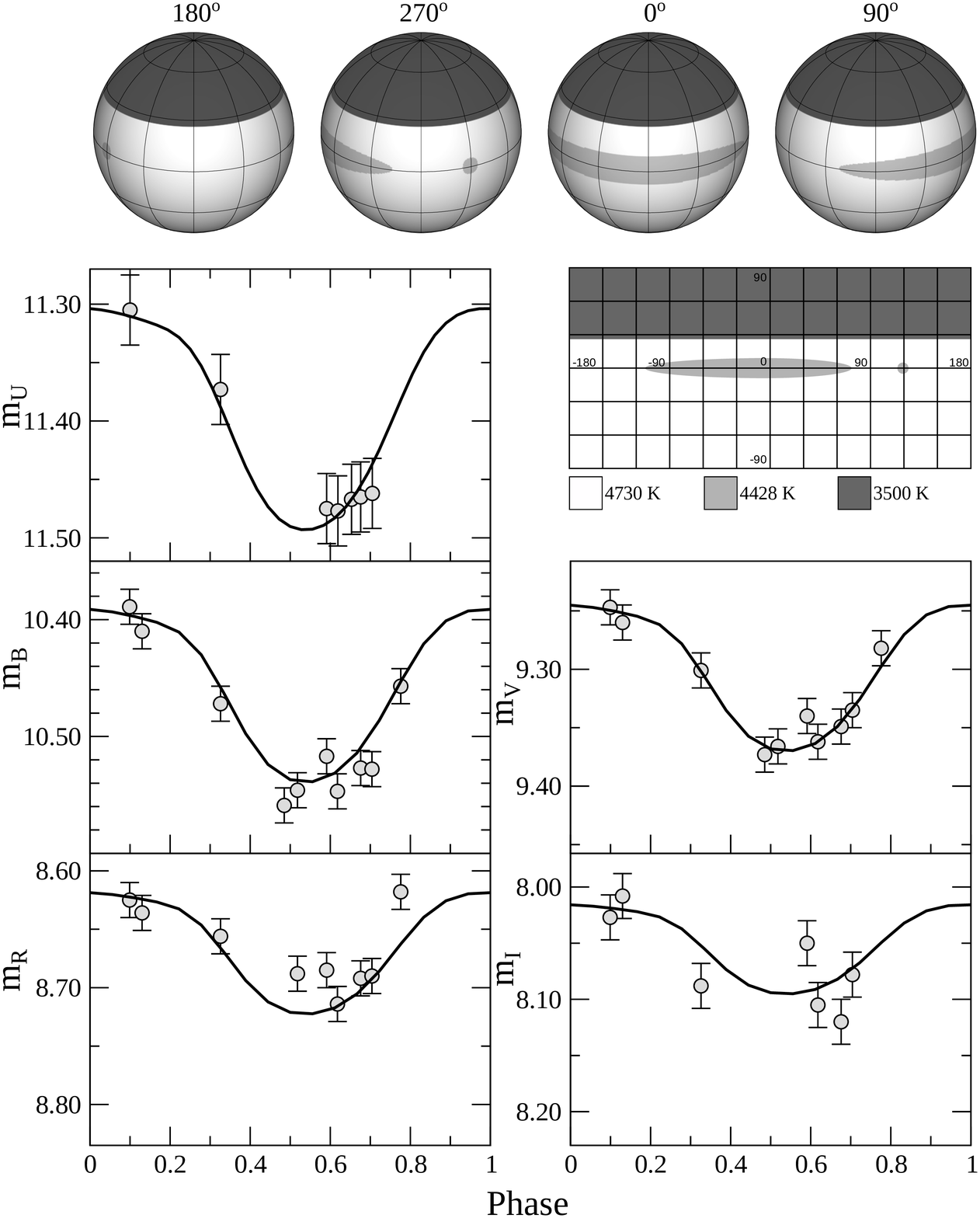}
\caption{Results of our modelling of the light curves and the surface of \pz\  from the observations at epoch 2 from November~22 to December~2, 2015, and from December~13, 2015, to February~6, 2016. The view of the surface at different phases is at the top. The surface map is in the middle, and the remaining panels show the observed and calculated light curves in different bands.}
\label{fig:synth_AA}
\end{figure}

The results of our calculations in the form of final model parameters are presented in Table~\ref{tab:param}. Figures~\ref{fig:synth_Crao} and~\ref{fig:synth_AA} for the observations at epochs~1 and~2 show the surface maps of \pz\ corresponding to the derived model parameters, the views of the stellar hemisphere at different phases (the longitude of the central meridian is indicated at the top), and the light curves of \pz\ calculated with the model used;
their accuracy is 0.007~mag for the $B$ and $V$ observations at epoch~1 and 0.009--0.013~m at epoch~2.

% Figure 6

% Figure 7
\clearpage

\section{DISCUSSION AND CONCLUSIONS}

The constructed surface models describe the currently most accurate observed light curves of \pz\ in different bands over 2015--2016. The observed light curves were modelled only in the $B$ and $V$ bands, which have a better absolute calibration. However, the calculated light curves in other bands also describe well both the shape and the amplitude of the observed variability.

Our model is also sufficient to describe all of the previous less accurate observations, primarily the longest ASAS-3 data series. We can explain the long-period brightness variations by a change in the sizes and temperature of the polar cool spot and the 34-day photometric period and its amplitude by the sizes and temperatures of the warm spots. The unspotted surface temperature \Teff=4730~K agrees well with the maximum brightness of \pz\ since 1899 in different bands and with the spectrum of an active giant \citep{2015MNRAS.446...56P}, while the temperatures of the warm spots turned out to be very close at different times of observations (see Table~\ref{tab:param}). 

The model in this paper is very rough, and it is
interesting to compare it with the surface maps of \rs\ stars obtained by Doppler tomography. The range of spot temperatures for \pz\ corresponds to their typical values for \rs\ stars \citep[see, e.g.][]{2000A&A...357..608S, 1998A&A...330.1029W, 2010A&A...515A..14K}. A constant cool polar region with
a temperature of $\sim$3500~K is present on the maps of many such stars \citep{2002A&A...389..202O, 2003A&A...402.1073G, 2000A&A...357..608S, 1998A&A...330.1029W, 2010A&A...515A..14K}. In these papers, however, it is noted that the polar region can be an artefact due to the limitations of the applied technique, because in the spectral line profile at zero relative radial velocity the line is identified with the core, which, first, may be described poorly by the model and, second, has a signal-to-noise ratio smaller than
that averaged over the spectrum. The only interferometric map of the \rs\ star $\zeta$~And available to date nevertheless clearly demonstrates the presence
of a cool polar region with a temperature of $\sim$3500--3600 K \citep{2016Natur.533..217R}. However, its radius is much smaller than that in our model and does not exceed 20\deg. Comparison of the surface maps for such stars shows that cool spots are present not only in the polar regions but also on the remaining stellar surface, occupying a significant fraction of the
visible hemisphere \citep{2010A&A...515A..14K}. However, the distribution of these spots is more or less uniform, so that their relative area on the visible hemisphere remains almost constant; as a consequence, there no modulation with the stellar rotation. Thus, our polar spot model reflects the total area of the cool spots. The sizes and shapes of the warm spots also reflect their distribution in longitude, whose maximum determines the active longitude. The active region located asymmetrically relative to the zero meridian. Whereas at the beginning of 2015 it appeared earlier, in 2016, on the contrary, it lagged behind this meridian. For this reason, the light curve at different epochs
can slightly disagree with the ephemeris, which also revealed by a periodogram analysis of the ASAS-3 photometric data. However, in view of the above limitations, we cannot take into account the differential stellar rotation that will also manifest itself in model as a shift of the spot maximum. 

Since the distribution of spots distorts the spectral line profiles, a change in the radial velocities determined from them is also expected:
$$
\overline{v_{rad}}=v_0 \, \frac{\int\textrm{d}I_V \, \textrm{sin}\,l \, \textrm{sin}\,\psi \, \textrm{sin}\,i}{\int\textrm{d}I_V}
$$
where $v_0$ = 11.4~\kms\ is the equatorial rotational velocity of \pz, $I_V$ is the intensity distribution over the stellar surface in $V$, and $i$ = 67\deg\ is the inclination of the rotation axis. Our calculations show maximum values of $v_{rad}$ no larger than 0.15~\kms\ which is comparable to the accuracy of determining the radial velocities of \pz\ and will not affect noticeably the result.

\section*{ACKNOWLEDGMENTS}

This work was supported in part by Program P-7 of the Presidium of the Russian Academy of Sciences and the Russian Foundation for Basic Research (project no. 15-02-06046).

\bibliographystyle{mnras}
\bibliography{paper}

\end{document}